\documentclass[aps, showpacs, preprintnumbers]{revtex4}
\begin{document}
\title{Quasinormal modes for the SdS black hole : an analytical 
approximation scheme}

\author{V. Suneeta \footnote{e-mail: suneeta@theorie.physik.uni-muenchen.de}}

\affiliation{Theoretische Physik, Ludwig-Maximilians-Universit\"{a}t,\\
Theresienstrasse 37, D-80333, M\"{u}nchen, Germany \\ }

\begin{abstract}
Quasinormal modes for scalar field perturbations of a 
Schwarzschild-de Sitter (SdS)
black hole are investigated. An analytical approximation is proposed for
the problem. The quasinormal modes are evaluated for this approximate 
model in the limit when black hole mass is much smaller than the radius
of curvature of the spacetime. The model mirrors some striking features
observed in numerical studies of time behaviour of scalar perturbations 
of the SdS black hole. In particular, it shows the presence of two sets of
modes relevant at two different time scales, 
proportional to the surface gravities of the black hole and
cosmological horizons respectively. 
These quasinormal modes are not complete - another
feature observed in the numerical studies. Refinements
of this model to yield more accurate quantitative agreement with numerical
results are discussed. Further investigations of this model are outlined,
which would provide a valuable insight into time behaviour of 
perturbations in the SdS spacetime.

\end{abstract}
\date{\today}
\maketitle
\section{Introduction}
A characteristic feature of the response of a black hole
to external perturbations is the appearance of quasinormal modes (QNMs). The
presence of these modes was first noted in a study of perturbations of the
Schwarzschild black hole, by Vishveshvara \cite{vish}. Since then, QNMs for
asymptotically flat black holes have been computed by a variety of numerical
and analytical approximation methods. A detailed review of this work can be
found in \cite{nollert, kokkotas}.

QNMs were first found in a stability analysis of black holes. However, the observation
that these damped oscillations are intrinsic characteristics of the black hole
exterior geometry and depend only on the black hole parameters implies that they
are the imprint of a black hole in its response to perturbations. Further, QNMs are 
even seen at intermediate/late times in fully non-linear 
situations like systems undergoing
gravitational collapse. Thus they are expected to play a significant role in the
search for gravitational waves and black holes. Recent evidence for a non-zero
positive cosmological constant points to the importance of studying black holes
in such a background. The simplest black hole in this class is a Schwarzschild
black hole in de Sitter space (SdS). The QNM spectra for gravitational 
perturbations of this black hole have been investigated in \cite{mellor, moss} using numerical
and analytical approximation techniques. QNM spectra for 
the SdS black hole in the near-extremal case when the two horizons
are nearly coincident (i.e, the black hole mass
and radius of curvature of the spacetime of the same order) 
are derived in \cite{lemos}. An 
interesting numerical study of evolution of scalar fields in the SdS spacetime
has been performed in \cite{bckl, bclp}. This numerical study reveals the differences
between the response of the Schwarzschild black hole and the SdS black hole to
perturbations in the physically interesting regime where the black hole
mass is much smaller than the radius of curvature of the spacetime (which is
a relevant regime because the
cosmological constant is small). The SdS black hole shows QNM behaviour similar to the 
Schwarzschild black hole at intermediate times, and soon changes
to a power law decay. At late 
times it again shows QNM behaviour (exponential decay) where the QNMs are 
now proportional to the surface gravity of the cosmological horizon.
For the Schwarzschild black hole, analytic approximations have been used to give
a general (approximate) formula for the QNMs as a function of the black hole 
parameters and the angular mode \cite{ferrari}.  
It is of interest to compute, in the same spirit, the
QNMs for the SdS black hole in an analytical approximation scheme in the physically
relevant regime.
The advantage is that there exist known numerical studies which 
reveal characteristic features in the time decay of
a field in the SdS black hole. The analytical approximation
could be refined to reproduce these features, and at the same time 
would give a general expression for the QNMs as a function of the spacetime parameters
and the angular mode. The approximate model, being simpler to study may also
be possible to analyse completely. It would thus give insight into the 
qualitative features of the complete time behaviour
of the field in SdS spacetime. Further, the near-extremal limit and its
effect on the QNMs and time behaviour of fields could be 
explored in this model.

In the recent past, the QNMs of Anti-de Sitter (AdS) black holes have been studied
extensively. A detailed numerical study   
of QNM decay of scalar fields in AdS black hole
backgrounds in various dimensions was performed in \cite{mann1}.
Exact computations of QNMs for the BTZ black hole in $(2+1)$ dimensions were demonstrated
first in
\cite{gs} and subsequently higher order modes (as also numerical results for gravitational
perturbations of SAdS black holes) are shown in \cite{lemos1}. More general
numerical computations of QNMs for higher dimensional AdS black holes are found in
\cite{hh}. Decay of scalar fields coupled to curvature in topological AdS black hole
backgrounds is studied numerically 
in \cite{mann2,mann3} and analytically in \cite{zan}. An interpretation
of the QNMs for an AdS black hole in the dual CFT using the AdS/CFT correspondence
is provided in \cite{bss} (see also \cite{son}). 
An interesting question is whether the QNMs of the SdS 
black hole have an interpretation in a dual CFT using the recently proposed 
dS/CFT correspondence \cite{strominger} (for a study of pure de Sitter QNMs in this
context, see \cite{branco}). The study of the time behaviour 
of perturbations in an analytical model for the SdS spacetime could be used
to study such a possible interpretation.

In this paper, we propose an analytical model to study the
time behaviour of perturbations in the SdS black hole. We compute the QNMs of 
this model in the regime when the black hole mass is small compared
to the radius of curvature of the spacetime. The SdS spacetime is bounded by
the black hole and cosmological horizon. Our analytical approximation consists 
of approximating the SdS potential by a potential that reproduces its 
asymptotic behaviour at both boundaries. In section 2, we enumerate the
QNM boundary conditions for the SdS black hole, and describe our approximation
potential along with the motivations for our choice. 
The chosen potential has a discontinuity in its second derivative at the
maximum. In section 3, we describe the exact solutions to this potential and list
the correct matching coefficients at the maximum of the potential. 
The actual computation of the QNMs is done in the regime of 
interest in the next section. Two sets of QNMs, one proportional to the black hole 
horizon surface gravity
and the other proportional to the surface gravity of the cosmological horizon
are found - the first set describing the time behaviour of the field
at certain intermediate times, and the latter describing late-time behaviour. 
A qualitative picture for the time scales where each of these two 
sets of QNMs comes into play is given - and it is discussed why this behaviour
is expected only in the regime where the black hole mass is small compared to 
radius of curvature. 
The real part of the QNMs depends on the maximum of the potential. 
Our approximation consists of identifying the value of the 
maximum of the model potential with
that of the SdS potential. We compute the maximum value of the SdS potential in 
section 5. This unfortunately is difficult to do for any general angular mode.
We therefore consider the small angular mode and large angular mode cases 
separately while computing the SdS maximum. 
In our last section, we present a detailed comparison of the results of this
model with numerical studies. There is good qualitative agreement in the most
striking feature of the numerical studies, the presence of the two sets of QNMs 
relevant at two different time scales, and which are 
proportional to the black hole and cosmological horizon surface gravities 
respectively.
We also show in this section, non-completeness of QNMs in our model - that is
also expected from the observation of power law behaviour at intermediate times
for the SdS black hole. We discuss how the model could be refined or modified 
to reproduce the more quantitative features of the numerical results like
the dependence of the QNMs on the angular mode. We also discuss the exciting
questions which can be addressed for this model, and would shed light
on our understanding of perturbations of the SdS spacetime.

\section{Analytical approximation of the SdS potential}

The metric  
for the Schwarzschild-de Sitter spacetime in four dimensions for   
cosmological constant $\Lambda={3}/{l^2}$ is given by
\begin{eqnarray}
ds^2 = -(N)^2~dt^2 + (N)^{-2}~dr^2 + r^2~(d\Omega)^2~,
\label{met3}
\end{eqnarray}
with
\begin{eqnarray}
N = \sqrt{(1~-~\frac{2 M}{r}~-~\frac{r^2}{l^2})}~.
\label{metdef3}
\end{eqnarray}
There are now two horizons given by the zeroes of $N^2$. These are the two real positive
roots of $N^2$, 
\begin{eqnarray}
r_{b}&=&\frac{2}{\sqrt{\Lambda}} 
\sin [\frac{1}{3} \sin^{-1} (3 M \sqrt{\Lambda})] ~~\mbox{and} \nonumber \\
r_{c}&=&\frac{2}{\sqrt{\Lambda}} 
\sin [\frac{1}{3} \sin^{-1} (3 M \sqrt{\Lambda}) ~+~\frac{2\pi}{3}]~.
\label{horizons}
\end{eqnarray}

The third root of $N^2$ is negative, and given by $r_{0}= -(r_{b}+r_{c})$.
$r_{b}$ is the black hole horizon and $r_{c}$ the cosmological horizon.
As in the asymptotically flat case, the spacetime has a curvature
singularity at $r=0$.

We consider a massless scalar field $\Phi$  in this background. 
Then, the Klein-Gordon equation for the field is
\begin{equation}
\partial_{\mu}(\sqrt{-g} g^{\mu \nu} \partial_{\nu})\Phi = 0~.
\end{equation}
Using an ansatz  
for the field 
\begin{eqnarray}
\Phi = \frac{1}{r}~ \chi(r)~ e^{i \omega t}~Y_{LM}(\theta, \phi)~,
\end{eqnarray}
and by going to the tortoise coordinate $x$ 
where $dx = {dr}/{N^2} $,  the Klein-Gordon equation is
\begin{eqnarray}
- \frac{d^2 \chi}{d x^2} + V_{L}(r) \chi = \omega^2 \chi~,
\label{kgeqn}
\end{eqnarray}
where
\begin{eqnarray}
V_{L}(r) =  (1- \frac{2 M}{r}- \frac{r^2}{l^2}) (\frac{2 M}{r^3} 
- \frac{2}{l^2} + \frac{L(L+1)}{r^2} )~.
\label{potential}
\end{eqnarray}
\noindent\\
When $r=r_{b}$, $x =-\infty$. When $r=r_{c}$, $x = \infty$. We
are interested in studying scalar field perturbations in the region
bounded by these two horizons. The potential $V_{L}(r)$ goes 
exponentially to zero as a function of the $x$ coordinate as one 
approaches the two horizons. However, the rate of approach is not the
same - it depends on the surface gravities associated with the 
horizons. The surface gravity at a horizon $r_{h}$ is defined as
$\alpha_{h}= \frac{1}{2} \left| {df}/{dr} \right | _{r_{h}}$.
Thus the surface gravities associated with the black hole
and cosmological horizons, $\alpha_{b}$ and $\alpha_{c}$ 
respectively are
\begin{eqnarray}
\alpha_{b} &=& \frac{(r_{c}-r_{b})(r_{b}-r_{0})}{2 l^2 r_{b}}~~\mbox{and}~~\\
\alpha_{c} &=& \frac{(r_{c}-r_{b})(r_{c}-r_{0})}{2 l^2 r_{c}}~. 
\label{surfaceg}
\end{eqnarray}
The potential $V_{L}(r)$ goes to zero as $e^{2 \alpha_{b} x}$ as
$x \rightarrow -\infty$, i.e as the black hole horizon is 
approached. 
When $x \rightarrow \infty$, i.e as the cosmological horizon is
approached, $V_{L}(r)$ goes to zero as $e^{-2 \alpha_{c} x}$.
In studying the time evolution of the perturbation, it is important
to first compute the QNMs for the problem. QNMs are given in this case
by solutions to (\ref{kgeqn}) with specific boundary conditions -
{\em ingoing} at the black hole horizon, and {\em outgoing} at the
cosmological horizon. These are the physically motivated boundary
conditions for the problem as a timelike observer in this spacetime
can send messages, but not receive them through the (future)
cosmological horizon. The boundary condition at the black hole horizon
is the usual one for computing QNMs of black holes and represents the
classical absorption by the black hole.
The QNM boundary conditions for the field are thus:
\begin{eqnarray}
\chi &\propto& \exp (i \omega x) ;~~ x \rightarrow - \infty ~~\mbox{and}
 \nonumber \\
\chi &\propto& \exp (-i \omega x) ;~~ x \rightarrow + \infty~. 
\label{qnmbc}
\end{eqnarray}

$\omega$ is now complex and $\omega~=~\omega_{0} + i \Gamma$, 
where $\Gamma > 0$.

Thus, the QNMs can be obtained by solving (\ref{kgeqn}) with the 
above boundary conditions. However, exact solutions to this equation 
are not known. In fact, such exact solutions are not available even
for the Schwarzschild black hole in an asymptotically flat
spacetime. Instead, what is done in the asymptotically flat case
is to either solve the equation numerically and compute the QNMs,
or to analytically approximate the Schwarzschild potential by an
exactly solvable potential.
This potential is described by parameters which can be used
to fit the potential to the Schwarzschild potential
\cite{ferrari}. We follow the second 
approach of an analytic approximation in computing the QNMs of the SdS black hole. 

In our search for a potential that would be a good analytic approximation
to the SdS potential, we set out to reproduce the behaviour of the
SdS potential as $x \rightarrow \pm \infty$. The approximate
potential $V$ that we have chosen has the following
features :

1. Its second derivative is discontinuous at $r=r_{m}$, where
$r_{m}$ is the value of $r$ for which the actual SdS potential
has a maximum, i.e a solution of ${dV_{L}}/{dr} = 0$.\\

2. In the region $r_{b} < r \leq r_{m}$, 
the potential is
\begin{eqnarray}
V_{1} = \frac{V_{m}}{[\cosh \alpha_{b}(x-x_{m})]^2}~.
\label{v1}
\end{eqnarray}
$x_{m}$ is the value of the $x$ coordinate when $r=r_{m}$.
$\alpha_{b}$ is given by (\ref{surfaceg}). 
$V_1$ is referred to as the Poschl-Teller potential in literature,
and was first used in \cite{ferrari} to approximate the Schwarzschild potential.
$V_m$ is taken to be the maximum of the SdS potential.

3. In the region $r_{m} \leq r < r_{c}$, the potential is taken to
be again the Poschl-Teller potential, but with a different choice of parameter.
It is
\begin{eqnarray}
V_{2} = \frac{V_m}{[\cosh \alpha_{c}(x-x_{m})]^2}~.
\label{v2}
\end{eqnarray}
$\alpha_{c}$ is given by (\ref{surfaceg}).

As can be checked, this potential reproduces the asymptotic form of the SdS potential at the
two boundaries. Furthermore, this potential is exactly solvable. Therefore what remains is a
computation of $V_{m}$, the value of the SdS potential at its maximum.  
Computing the maximum of the SdS potential is difficult as the potential is
complicated. We are forced
to work in the regime $M\ll l$.  
The Poschl-Teller has been used to approximate the SdS potential recently \cite{moss}.
However, in this case, {\em one} Poschl-Teller potential
is used throughout - and this is a good approximation only for the nearly extreme SdS black 
hole, where the two horizons are very close (i.e, $M$ and $l$ are of the same order) \cite{lemos}. 
In the other regime $M\ll l$, one Poschl-Teller potential 
provides a poor approximation
to the SdS potential as  
the asymptotic forms of the SdS potential at both the boundaries cannot be reproduced
by it. This problem is solved by our choice for the potential,
$V$. It would be exciting if the near-extremal limit in our model led to the QNMs
found in \cite{lemos}.
We must now check what features, qualitative or quantitative, we reproduce of numerical
studies of dynamics of a scalar field in the SdS spacetime done in \cite{bckl, bclp} by a 
study of this potential. 

We now proceed to compute the QNMs of the approximate potential $V$ which is given by $V_{1}$ for
$r \leq r_{m}$ and $V_{2}$ for $r \geq r_{m}$.
We are interested in the solutions to the equation :
\begin{eqnarray}
- \frac{d^2 \chi}{d x^2}~ +~ V \chi ~=~ \omega^2 \chi~.
\label{appeqn}
\end{eqnarray}
For both $V_1$ and $V_2$, exact solutions can be found.
We first examine the exact solutions to (\ref{appeqn}) with both $V = V_{1}$ and $V = V_{2}$. Then,
the solutions have to be matched at $r=r_{m}$. Finally, we must pick the solutions that obey the QNM
boundary conditions at both the boundaries. 

\section{Exact solutions for the potential}
Let us consider the solutions to (\ref{appeqn}) with $V=V_1$. The solutions to this (Poschl-Teller)
potential are known. They can be written in terms of the hypergeometric functions after a change of
variables to $\xi$ where ${\xi}/({1- \xi}) ~=~e^{2\alpha_{b} (x-x_{m})}$. 
Then, two linearly independent solutions are :
\begin{eqnarray}
\chi_{1} &=& [\xi (1-\xi)]^{\frac{i\omega}{2 \alpha_{b}}} F(a, b, c; \xi)~~\mbox{and} \\
\label{s1}
\chi_{2} &=& [\xi (1-\xi)]^{\frac{i\omega}{2 \alpha_{b}}} (\xi)^{1-c} F(a-c+1, b-c+1, 2-c; \xi)~.  
\label{s2}
\end{eqnarray}
We have
\begin{eqnarray}
a = \frac{1}{2} + \sqrt{\frac{1}{4} - \frac{V_{m}}{\alpha_{b}^{2}}} + 
\frac{i\omega}{\alpha_b}~,~
b = \frac{1}{2} - \sqrt{\frac{1}{4} - \frac{V_{m}}{\alpha_{b}^{2}}} + 
\frac{i\omega}{\alpha_b}~,~ 
c = 1 + \frac{i\omega}{\alpha_{b}}~.
\label{defn1}
\end{eqnarray}
The black hole horizon is at $\xi = 0$. At the maximum of the potential, when $x=x_{m}$, $\xi={1}/{2}$.

Solutions to (\ref{appeqn}) with $V=V_2$ are similar to the previous case, but with $\alpha_{b}$
replaced by $\alpha_{c}$. 

Define $\xi_{1}$ where ${\xi_{1}}/{1- \xi_{1}}=e^{2\alpha_{c} (x-x_{m})}$. 
\begin{eqnarray}
\tilde \chi_{1} &=& [\xi_{1} (1-\xi_{1})]^{\frac{i\omega}{2 \alpha_{c}}} F(a_{1}, b_{1}, c_{1}; \xi_{1})~~\mbox{and}\\
\label{st1}
\tilde \chi_{2} &=& [\xi_{1} (1-\xi_{1})]^{\frac{i\omega}{2 \alpha_{c}}} (\xi_{1})^{1-c_{1}} 
F(a_{1}-c_{1} +1, b_{1}-c_{1} +1, 2-c_{1}; \xi_{1})~.  
\label{st2}
\end{eqnarray}
Here
\begin{eqnarray}
a_{1} = \frac{1}{2} + \sqrt{\frac{1}{4} - \frac{V_{m}}{\alpha_{c}^{2}}} + 
\frac{i\omega}{\alpha_{c}}~,~
b_{1} = \frac{1}{2} - \sqrt{\frac{1}{4} - \frac{V_{m}}{\alpha_{c}^{2}}} + 
\frac{i\omega}{\alpha_{c}}~,~ 
c_{1} = 1 + \frac{i\omega}{\alpha_{c}}~.
\label{defn2}
\end{eqnarray}
The next step is to match an arbitrary superposition of the two solutions (\ref{s1}) and (\ref{s2}) with
an arbitrary superposition of (\ref{st1}) and (\ref{st2}) - at $x = x_{m}$. This involves matching both
the wave functions and their first derivatives across $x = x_{m}$. Let us use the following notation:
$\chi_{1}(x_{m}) = X$, ~~$\chi_{2}(x_{m}) = Y$. Also, ~~$\frac{d\chi_{1}}{dx} (x_{m}) = X', 
~~\frac{d\chi_{2}}{dx} (x_{m}) = Y'$. ~~$\tilde \chi_{1}(x_{m}) = \tilde X , ~~\tilde \chi_{2}(x_{m}) = \tilde Y$,
and ~~$ \frac{d\tilde \chi_{1}}{dx} (x_{m}) = \tilde X',
~~\frac{d\tilde \chi_{2}}{dx} (x_{m}) = \tilde Y'$. 

\noindent\\
Then matching the wave functions and their derivatives across $x=x_{m}$ yields
\begin{eqnarray}
A X + B Y = \tilde A \tilde X + \tilde B \tilde Y ~;~
A X' + B Y' = \tilde A \tilde X' + \tilde B \tilde Y'~.
\label{match}
\end{eqnarray}
\noindent\\
Here, $A$, $B$, $\tilde A$ and $\tilde B$ are arbitrary coefficients of superposition 
of the two
linearly independent solutions, which are then 
$\chi = A \chi_{1} + B \chi_{2}$ for $r < r_{m}$ and
$\chi = \tilde A \tilde \chi_{1} + \tilde B \tilde \chi_{2}$ for
$r > r_{m}$. 

\noindent\\
$A$ and $B$ are related to $\tilde A$ and $\tilde B$ as :

\begin{eqnarray}
\tilde A = A~\frac{(X \tilde Y' - X' \tilde Y)} 
{(\tilde X \tilde Y' -  \tilde X' \tilde Y)}
~~+~~B~\frac{(Y \tilde Y' - Y' \tilde Y)}
{(\tilde X \tilde Y' - \tilde X' \tilde Y)}~, 
\label{tildea}
\end{eqnarray}

\begin{eqnarray}
\tilde B = A~\frac{(X \tilde X' - X' \tilde X)}
{(\tilde Y \tilde X' - \tilde Y' \tilde X)}
~~+~~B~\frac{(Y \tilde X' - Y' \tilde X)}
{(\tilde Y \tilde X' - \tilde Y' \tilde X)}~. 
\label{tildeb}
\end{eqnarray}
\noindent\\
We recognise that at $x=x_{m}$, the value of $\xi$ that appears in
(\ref{s1}) and (\ref{s2}) is ${1}/{2}$. 
Using this and properties of the 
hypergeometric functions, we can write :
\begin{eqnarray}
X &=&\sqrt{\pi} \exp (- \frac{i\omega \ln 2}{\alpha_{b}})
\frac{\Gamma(\frac{1}{2} + \frac{a}{2} + \frac{b}{2})}
{\Gamma(\frac{1}{2} + \frac{a}{2})\Gamma(\frac{1}{2} + \frac{b}{2})}~,
\label{x} \\
X' &=& \frac{\sqrt{\pi}\alpha_{b}}{2} \exp (- \frac{i\omega \ln 2}{\alpha_{b}}) \frac{a~b}{c}
\frac{\Gamma(2 + \frac{i\omega}{\alpha_b})} 
{\Gamma(1 + \frac{a}{2})\Gamma(1 + \frac{b}{2})}~, 
\label{x'} \\
Y &=& \sqrt{\pi}\exp (\frac{i\omega \ln 2}{\alpha_{b}}) 
\frac{\Gamma(1 - \frac{i\omega}{\alpha_b})}
{\Gamma(1 - \frac{a}{2})\Gamma(1 - \frac{b}{2})}~,
\label{y} \\
Y' &=& -i\omega\sqrt{\pi} \exp (\frac{i\omega \ln 2}{\alpha_{b}}) 
\frac{\Gamma(1 - \frac{i\omega}{\alpha_b})}
{\Gamma(1 - \frac{a}{2})\Gamma(1 - \frac{b}{2})} \nonumber \\
&+& \frac{\alpha_b}{2}
\frac{(a-c+1)(b-c+1)}{2-c} F(a-c+2, b-c+2, 3-c; \frac{1}{2})~.
\label{y'}
\end{eqnarray}
\noindent\\
$\tilde X, \tilde Y, \tilde X'$ and $\tilde Y'$ are given by the same expressions as above,
except that we replace $\alpha_{b}$ with $\alpha_{c}$. Consequently $a$, $b$ and $c$
are also replaced by their counterparts $a_{1}$, $b_{1}$ and $c_{1}$. We therefore
do not display those expressions here.

\section{ Computation of quasinormal modes}

We search for solutions to the problem satisfying quasinormal mode boundary conditions
at the black hole and cosmological horizons given by (\ref{qnmbc}).   
The solution obeying the QNM boundary conditions at the black hole horizon
$\chi_{qnm}$ for $x > x_{m}$ is given by $\tilde A \tilde \chi_{1} +
\tilde B \tilde \chi_{2}$. $\tilde A$ and $\tilde B$ for $\chi_{qnm}$ are given by
(\ref{tildea}) and (\ref{tildeb}) with $B=0$. Near the cosmological horizon, i.e
as $x \rightarrow \infty$, this solution is
\begin{eqnarray}
\chi_{qnm} \sim \tilde A P_{1} \exp (-i\omega x) + \tilde A Q_{1} \exp (i\omega x)
+ \tilde B P_{2} \exp (-i\omega x) + \tilde B Q_{2} \exp (i\omega x)~,
\label{asyqnm}
\end{eqnarray}
where 
\begin{eqnarray}
P_{1} &=& \frac{\Gamma (c_{1}) \Gamma (c_{1} - a_{1} - b_{1})}{\Gamma (c_{1} - a_{1})
\Gamma (c_{1} - b_{1})}~,
\label{p1} \\
Q_{1} &=& \frac{\Gamma (c_{1}) \Gamma (a_{1} + b_{1} - c_{1})}{\Gamma (a_{1}) \Gamma (b_1)}~,
\label{q1} \\
P_{2} &=& \frac{\Gamma (2 - c_1) \Gamma (c_1 - a_1 - b_1)}{\Gamma (1 - a_1) 
\Gamma (1 - b_1)}~,
\label{p2} \\
Q_{2} &=& \frac{\Gamma (2 - c_1) \Gamma (a_1 + b_1 - c_1)}{\Gamma (a_1 - c_1 + 1)
\Gamma (b_1 - c_1 + 1)}~.
\label{q2}
\end{eqnarray}
\noindent\\
For (\ref{asyqnm}) to obey QNM boundary conditions at the cosmological horizon given
by (\ref{qnmbc}), we must have
\begin{eqnarray}
\tilde A Q_{1} + \tilde B Q_{2} = 0~,
\label{qnmeqn}
\end{eqnarray}
\noindent\\
Substituting for $\tilde A$ and $\tilde B$ from (\ref{tildea}) and (\ref{tildeb}) with
$B=0$, the above equation is
\begin{eqnarray}
(X \tilde Y' - X'\tilde Y) \frac{\Gamma (c_1)}{\Gamma (a_1) \Gamma (b_1)} ~~-~~
(X \tilde X' - X' \tilde X) \frac{\Gamma (2 - c_1)}{\Gamma (a_1 - c_1 + 1)
\Gamma (b_1 - c_1 + 1)} = 0~.
\label{qnmeqn1}
\end{eqnarray}
The special frequencies $\omega$ that solve the above equation are the quasinormal
mode frequencies for this problem. However, as can be seen from the expressions for
$X$ and $X'$ from (\ref{x}) and (\ref{x'}) - the zeroes of $X$ and $X'$ do not occur
for the same set of frequencies. The zeroes of the above equation cannot therefore
be computed simply. However, here we reflect on what we must {\em expect}. From the 
numerical work in \cite{bckl, bclp}, we note the presence of two different patterns
of exponential decay of the field with time. For intermediate times, the field decays
exponentially with the exponent proportional to the black hole surface gravity. This
is a typical QNM time behaviour where the QNM frequency is proportional to the black
hole surface gravity. This temporal behaviour is similar to that of a field in a 
Schwarzschild black hole background. This exponential decay is followed by a power law decay 
of the field - but for late times, there is a switch back to exponential decay. Now
the exponent is proportional to the surface gravity of the cosmological horizon. This
exponential decay is again a QNM time behaviour, but the QNMs are now proportional
to the surface gravity of the cosmological horizon. This numerical result was seen
for a choice of {\em well separated} length scales $M$ and $l$.

We expect a similar behaviour (i.e two sets of QNMs) for our approximate potential $V$.
At very early times, there is a direct propagation of the perturbation without
scattering by the potential.
At later times, the perturbation gets multiply scattered by
the potential. It therefore starts reflecting the properties of the potential.
The time behaviour of the field here is dominated by the QNMs.
In our case, the second derivative of the potential has a discontinuity at $r=r_{m}$. The QNMs
which we have obtained (and which are supposed to describe the time 
evolution after the very early times) also reflect this.
When the perturbation gets scattered for $r_{b}< r \leq r_{m}$
by the potential (which is now $V_{1}$), its time evolution will be governed
by QNMs related to $V_{1}$ which should be proportional to $\alpha_{b}$ 
(as is typical for QNMs of a Poschl Teller potential). A rough estimate for the times
when this occurs is the light crossing time to traverse a distance of the
order of $r_{m}$, i.e ${r_{m}}/{c}$. When the perturbation travels a distance
much greater than $r_{m}$, it starts getting scattered by the potential $V_{2}$,
and its time evolution is governed by QNMs related to $V_{2}$, which we expect to
be proportional to $\alpha_{c}$. This roughly occurs at times of the order of 
the light crossing time
to traverse a distance $l$, i.e ${l}/{c}$. At such times and later,
the decay of the perturbation is governed solely by the QNMs related to $V_2$. 
However, the above statements are possible only when there is a clear separation of these
two interesting time scales. As we see in the next section,
if we have a clear separation of the length scales $M$ and $l$, this implies 
that $r_m \sim O(M)$. Thus a clear separation of the two interesting time scales 
$\frac{r_m}{c}$ and $\frac{l}{c}$ is then related to a clear separation of 
$\alpha_{c} \sim 1/l$ and $\alpha_{b} \sim 1/M$. We note that when such a clear separation
of the scales does not exist (as for e.g in the nearly extreme SdS black hole studied
in \cite{moss, lemos}) then we do not expect two different sets of QNMs.

Let us return to (\ref{qnmeqn1}). Since we have not yet specialised to the case where
the length scales are separated, we do expect the general solutions of this equation
for $\omega$ to be complicated. However, if we specialise to the case where 
$M \ll l$,
and work in an approximation where we can neglect $O(M/l)$ terms and higher, we
must see solutions $\omega$ that split into two sets as argued above. We therefore
proceed to study the various terms in (\ref{qnmeqn1}) and drop terms that can be
neglected in this approximation. 
For $M \ll l$, $\alpha_b \sim 1/M$ and $\alpha_c \sim 1/l$.
Solutions $\omega$ for the case of the two length scales being well-separated are of
the form $\omega \sim O(1/M)$ or $\omega \sim O(1/l)$.

We now look for solutions of the form $\omega \sim O(1/M)$. Then, in (\ref{qnmeqn1}),
the first term has a factor $(X\tilde Y' - X'\tilde Y)$. From the detailed 
expression for this factor, it is clear that this factor is proportional to
$\exp (i\omega l \ln 2)$. Now, $\omega~=~\omega_{0} + i\Gamma$ where $\Gamma > 0$.
Furthermore, $\Gamma \sim O(1/M)$. Therefore, this factor 
$(X\tilde Y' - X' \tilde Y)$ is proportional to $\exp (-l/M)$. On the other hand, 
the second term in (\ref{qnmeqn1}) has a factor $(X\tilde X' - X' \tilde X)$ which
is proportional to $\exp (l/M)$. Since $M\ll l$, the first term can be neglected
compared to the second term (Here, we have also used the asymptotic expansion of
the hypergeometric function for large parameter).
Therefore, in this approximation we must solve the 
equation
\begin{eqnarray}
X \tilde X' - X' \tilde X = 0~.
\label{app1}
\end{eqnarray}
Here again, we look at the detailed expressions for $X'$ in (\ref{x'}) and 
$\tilde X'$. Using these detailed expressions and multiplying (\ref{app1}) by 
$M$, we see that the first term 
in (\ref{app1}) $X\tilde X'$ is $O(M/l)$ compared to the second
term $X' \tilde X$. 
Neglecting the first term, we see that we must now look for zeroes of the
second term which lead to frequencies $\omega \sim O(1/M)$. These are given
by the zeroes of $X'$, which occur (as can be seen
on inspecting (\ref{x'}) when $1 + {a}/{2} = - n$ and $n \geq 0$ is
an integer.
Thus, the frequencies $\omega$ are :
\begin{eqnarray}
\omega ~=~ \alpha_{b} [\sqrt{\frac{V_{m}}{\alpha_{b}^{2}} - \frac{1}{4}} + i (2 n +
\frac{5}{2} ) ]~.
\label{omegaset1}
\end{eqnarray} 
We note, that above, we do not consider zeroes of $\tilde X$ as they lead to 
frequencies not of $O(1/M)$ as assumed while arriving at (\ref{app1}). 
We have nevertheless 
shown that in the $M\ll l$ approximation, there indeed exist solutions
$\omega \sim O(1/M)$ to (\ref{qnmeqn1}). 

We now consider the other possibility,
i.e $\omega \sim O(1/l)$. Now we can no longer neglect the first term of 
(\ref{qnmeqn1}) compared to the second term, as we did before. However, we multiply
the whole equation by $M$, and observe that part of the second term can still be
neglected as in the previous case of $\omega \sim O(1/M)$. More precisely,
the second term has a factor $(X \tilde X' - X' \tilde X)$ where the first term
is $O(M/l)$ compared to the second, and can be dropped. We now concentrate on the
approximate equation
\begin{eqnarray}
(X \tilde Y' - X'\tilde Y) \frac{\Gamma (c_1)}{\Gamma (a_1) \Gamma (b_1)} ~~+~~
X' \tilde X \frac{\Gamma (2 - c_1)}{\Gamma (a_1 - c_1 + 1)
\Gamma (b_1 - c_1 + 1)} = 0~.
\label{app2}
\end{eqnarray}
There are more $O(M/l)$ terms left in the above equation. However, we observe
that the first term has a factor $1/\Gamma (a_1)$. Thus zeroes of the above
equation where $\omega \sim O(1/l)$ could occur at poles of $\Gamma (a_1)$
which are also zeroes of $\tilde X$. We check for such solutions.
The poles of $\Gamma (a_1)$ occur 
when $\Gamma (a_1) = - n'$ and $n' \geq 0$ is an integer.
The corresponding frequencies are
\begin{eqnarray}
\omega = \alpha_{c}~ [ \sqrt{\frac{V_m}{\alpha_{c}^{2}} - \frac{1}{4}}~+~ i (n' + 
\frac{1}{2}) ]~. 
\label{polea1}
\end{eqnarray}
The zeroes of $\tilde X$ occur for ${a_1}+ {1} = - 2n$ where
$n \geq 0$ is an integer. The corresponding frequencies are
\begin{eqnarray}
\omega = \alpha_{c}~ [ \sqrt{\frac{V_m}{\alpha_{c}^{2}} - \frac{1}{4}}~+~ 
i (2 n + 
\frac{3}{2}) ]~. 
\label{omegaset2}
\end{eqnarray}
From (\ref{polea1}) and (\ref{omegaset2}), we see that the poles of 
$\Gamma (a_1)$
which are also zeroes of $\tilde X$ occur when, $n'$ is an odd integer.
These frequencies $\omega \sim O(1/l)$ are then the solutions of (\ref{app2}) and
in the $M\ll l$ approximation, the solutions of (\ref{qnmeqn1}).   

\section{Maximum of the SdS potential}

In the previous section, we argued for the presence of two sets of QNMs, one 
proportional to $\alpha_b$ and the other to $\alpha_c$, in the approximation
$M\ll l$. We also computed these frequencies which are given by 
(\ref{omegaset1}) and (\ref{omegaset2}). But it remains to evaluate $V_{m}$, the
maximum of the SdS potential $V_{L} (r)$. The maximum is a solution of the
equation ${dV_L}/{dr} = 0$, which is a sextic equation in $r$. For high
values of the angular mode number $L$, the potential is approximately
\begin{eqnarray}
V_{L}(r) ~\sim~  (1- \frac{2 M}{r}- \frac{r^2}{l^2})  
\frac{L(L+1)}{r^2}~. 
\label{largeLpot}
\end{eqnarray}
Then the maximum occurs at $r_m = 3M$. Thus, for large L, we can take
$V_{m} =V_L (3M)$. 
\begin{eqnarray}
V_L (3 M) = (\frac{1}{3} - \frac{9 M^2}{l^2})(\frac{2}{27 M^2} - \frac{2}{l^2}
+ \frac{L(L+1)}{9 M^2})~.
\label{vmlargel}
\end{eqnarray}
\noindent\\
Let us now look at small values of $L$. We first consider the case $L=0$.
For $L=0$, numerically, with $M=1$ and $\Lambda = 10^{-4}$, we can 
estimate the maximum, which is $r_m = 2.67$. 
For $M=2$ and the same value of $\Lambda$, $r_m =5.33$. This suggests that
$r_m =2.67 M$.
Analytically, the equation
to be solved is ${dV_0}/{dr} =0$, and the equation is
\begin{eqnarray}
2 \frac{r^6}{l^6} - \frac{M r^3}{l^4} - \frac{3 M r}{l^2} + \frac{8 M^2}{l^2} =0~.
\label{rzeromax}
\end{eqnarray}
Now, let us assume that the solution $r_{m} \sim O(M)$ as suggested by our numerical
results. Then, in the equation above, the first and second terms are $O( (M/l)^2)$
or higher powers of $(M/l)$ compared to the last two terms. Therefore, dropping these terms
we see that $r_m =\frac{8}{3} M$. This agrees with our numerical results. For the
QNMs with $L=0$, we can therefore set $V_m = V_{0} (\frac{8}{3} M)$.
\begin{eqnarray}
V_{0} (\frac{8}{3} M)= (\frac{1}{4} - \frac{64 M^2}{9 l^2})(\frac{27}{256 M^2} -
\frac{2}{l^2})~.
\label{vml0}
\end{eqnarray}
\noindent\\
For $L=1$, we can again solve the equation ${dV_1}/{dr}=0$, assuming as before
that the solution $r_m \sim O(M)$ and in the approximation $M\ll l$. We find that
$r_{m}= \frac{1}{4} (3 + \sqrt{73}) M$, i.e $r_m \sim 2.886 M$. 
We expect $r_m =3 M$ to be a good approximation to the maximum
for higher values of $L$. 

In all the above cases, we can explicitly check that ${V_{m}}/{\alpha_{b}^{2}}$
and ${V_m}/{\alpha_{c}^{2}}$ are greater than ${1}/{4}$. Thus, it is the real
part of the QNM frequencies that depends on $V_{m}$. Substituting for the value
of $V_m$ from (\ref{vmlargel}) in the expressions for the QNMs (\ref{omegaset1})
and (\ref{omegaset2}) - we get the explicit expression for the QNM frequencies for
$L > 0$. For $L=0$, using (\ref{vml0}) for $V_m$, we obtain more precise 
values for the QNM frequencies.

\section{Discussion}

In the previous sections, we have done an analysis of a perturbation in the potential
$V$ and given a sketch of its behaviour in time. We have found the presence of 
two sets of QNMs which are relevant at two different time scales (as discussed
in Section IV):

\begin{eqnarray}
\omega &=& \alpha_{b} [\sqrt{\frac{V_{m}}{\alpha_{b}^{2}} - \frac{1}{4}} + i (2 n +
\frac{5}{2} ) ] ~~\mbox{and}~~\\
\omega &=& \alpha_{c}~ [ \sqrt{\frac{V_m}{\alpha_{c}^{2}} - \frac{1}{4}}~+~ 
i (2 n + 
\frac{3}{2}) ]~. 
\end{eqnarray}

$V_m$ is determined as discussed in the previous section. We have also given
a qualitative picture for the times at which each of these sets of QNMs would
be relevant for the temporal behaviour of the scalar field.

We now address the question of how
well this reflects the time behaviour of a perturbation in the SdS potential. Numerical
studies have been done for scalar fields in the SdS spacetime \cite{bckl, bclp}. We 
indeed reproduce the most striking feature of the studies, the presence of two sets
of modes which are proportional to the surface gravities of the black hole and
cosmological horizons relevant at intermediate and late times respectively. 

Now let us make more quantitative comparison with the numerical results.
We begin with the intermediate times at which the first set of modes
(\ref{omegaset1}) are relevant. In the numerical work \cite{bclp}, a scalar 
field wave function with a Gaussian profile is chosen. Its subsequent time
evolution in the SdS spacetime is found by numerically integrating the Klein-Gordon
equation. A choice of $r_b =1$ and $r_c = 2000$ (such that we are in the regime
$M \ll l$) is taken. A graph of the field behaviour in an SdS spacetime versus
similar behaviour in a Schwarzschild spacetime is plotted for $L=0$ and
$L=1$. It is seen that for 
early times, the two graphs follow each other closely and display QNM oscillations - 
they start to
differ from each other
only for times much greater than $\frac{r_b}{c}$. 

Let us compare our first set of modes (\ref{omegaset1}) with the Schwarzschild QNMs which
seem to describe the behaviour in SdS well at these times. To do so, we can use 
an analytical approximation for the Schwarzschild QNMs by Ferrari and Mashhoon
\cite{ferrari} which approximates the QNMs well when their imaginary parts are not
too large. Then in this approximation, the real ($\omega^{S}_{R}$)
and imaginary ($\omega^{S}_{Im}$) parts
of the Schwarzschild QNMs are 
$\omega^{S}_{R} = \alpha (\sqrt{\frac{U_{0}}{\alpha^2} - \frac{1}{4}})$
and $\omega^{S}_{Im} = \alpha (n + \frac{1}{2})$. Here, $\alpha$ and $U_{0}$ are
given by the height and curvature of the Schwarzschild potential at its maximum
and $n$ is an integer.
$\alpha^{-1} = 3 \sqrt{3} M$. 
Comparing with the modes (\ref{omegaset1}), we see that for $M \ll l$,
$U_{0} \sim V_m$. Also $\alpha_b^{-1} \sim 4 M$ and therefore 
$\alpha_b \sim (1.3) \alpha$.
Thus we see that the real parts of the Schwarzschild QNMs and (\ref{omegaset1})
have approximately the same $M$ dependence. The angular mode $L$ dependence is also the
same in both real parts. However, coming to the imaginary parts, we see that our
set of QNMs reproduces only those Schwarzschild QNMs starting from $n=2$ and 
subsequently reproduces only alternate $n$ modes, the next one being $n=4$.
This probably has to do with our matching conditions
at $r_{m}$.

We proceed to compare the time behaviour of the field in this model with numerical
results for later times. In the next sub-section, we address intermediate time 
power law decay. The following sub-section compares the QNM behaviour 
at late times, i.e given by (\ref{omegaset2}) in our model with numerical results
- particularly with reference to angular mode dependence.

\subsection{Incompleteness of QNMs}
An important point in the numerical results \cite{bckl, bclp} is the observation
of a power law behaviour in time for the field at certain intermediate times, rather
than QNM behaviour. It is seen that the time decay of the field first displays
Schwarzschild QNM behaviour, then Schwarzschild power law decay. At later times, the 
power law decay becomes faster (to be followed eventually by exponential decay).

We examine whether there is a possibility of such interesting behaviour in our
model.
A detailed analysis has been done in \cite{ching} on the 
conditions (and types of potentials) for which both QNM and power law behaviour 
in time may be present at different times. Also, the more general question of 
the potentials for which the QNMs form a complete set has been addressed. The results
of the analysis are best expressed in terms of the Fourier transform of the Green's 
function for the system. This can be compactly written in terms of two auxiliary
functions $f(\omega, x)$ and $g(\omega, x)$ which are solutions to the homogenous
time-independent Klein Gordon equation, where $f$ satisfies the QNM boundary 
condition at one boundary (in our case, the black hole horizon) and $g$ satisfies
the QNM boundary condition at the other boundary (in our case, the cosmological
horizon). Then the Green's function is 
\begin{eqnarray}
\tilde G (x,y;\omega)&=& f(\omega,x) g(\omega,y)/W(\omega) ~~\mbox{for}~~ 0<x<y~, \nonumber \\
&=& f(\omega,y) g(\omega,x)/W(\omega) ~~\mbox{for}~~ 0<y<x~.
\label{greensfunction}
\end{eqnarray}

Here, the Wronskian $W(\omega) = g(\omega,x)f'(\omega,x) - f(\omega,x)g'(\omega,x)$
is independent of $x$. The QNMs are given by the zeroes of the Wronskian, i.e when the
functions $f$ and $g$ are linearly dependent. Following \cite{ching}, the QNMs do 
not describe the intermediate or late time behaviour completely when the functions
$g$ and $f$ are not analytic w.r.t $\omega$. Rather, non-analyticity of these functions
could lead to power law behaviour in time. $f$ and $g$ are analytic in $\omega$ if
the potential for the system $V$ has `no tail' at each of the boundaries
$x= -\infty$ and $x= +\infty$ in the following sense (condition for boundary at
$x=\infty$) :
\begin{eqnarray}
\int_{0}^{\infty} dx~ x |V(x)| < \infty ~,~~ 
\int_{0}^{\infty} dx~x e^{\alpha x} |V(x)| < \infty ~~\mbox{for any}~~ \alpha > 0~.
\label{potcond}
\end{eqnarray}
There is a corresponding condition for the boundary at $x=-\infty$. If this condition 
is violated for some $\alpha > \alpha_{0} > 0$, then $f$ and $g$ may not be 
analytic in $\omega$ for $\textrm{Im} \omega > \alpha_{0}$. 
Let us apply these results to our problem. For our problem, the function $f(\omega, x)$
obeying the QNM boundary condition at the black hole horizon is $\chi_{1}$, and 
for $x \rightarrow \infty$, it is given by the expression for 
$\chi_{qnm}$ in (\ref{asyqnm}). The function $g(\omega, x)$ obeying the QNM
boundary condition at the cosmological horizon is 
\begin{eqnarray}
g(\omega,x) = \frac{1}{(Q_{2}P_{1} - Q_{1}P_{2})} ~~(Q_{2} \tilde \chi_{1} - Q_{1} 
\tilde \chi_{2} )~.
\label{g}
\end{eqnarray}
It can be easily checked that the Wronskian $W(\omega)$ is 
\begin{eqnarray}
W(\omega) = -2 i \omega (\tilde A Q_{1} + \tilde B Q_{2})~.
\label{wronskian}
\end{eqnarray}
QNMs are frequencies for which the Wronskian is zero, 
and as expected, are given by solutions
to (\ref{qnmeqn}) which we discussed extensively in the previous sections. 

We are interested in the question of whether $f$ and $g$ are analytic in 
$\omega$. The potential $V$ for our problem goes to zero as $e^{-2 \alpha_{c} x}$
as $x \rightarrow \infty$ and as $e^{2 \alpha_{b} x}$ as $x \rightarrow -\infty$.
Therefore, conditions (\ref{potcond}) are violated for $\alpha > \alpha_{b}$ 
(coming from the condition at the black hole horizon) and for $\alpha > \alpha_{c}$
(coming from the condition at the cosmological horizon). We expect non-analyticity
for two sets of $\omega$ : $\textrm{Im} \omega > \alpha_{b}$ and 
$\textrm{Im} \omega > \alpha_{c}$. 

Let us now examine $f$ and $g$ to see if this is indeed the case. $f = \chi_{1}$.
Using the properties of hypergeometric functions for the specific values of
parameters $a$, $b$ and $c$ for the problem, we can write $f$ in terms of
the associated Legendre polynomial as
\begin{eqnarray}
\chi_{1} = [\xi (1-\xi)]^{\frac{i\omega}{2 \alpha_{b}}}
\Gamma (c) (\xi - \xi^{2})^{(1-a-b)/4} P_{(a-b-1)/2}^{(1-a-b)/2} (1 - 2\xi)~. 
\label{f}
\end{eqnarray}
Therefore $f$ is not analytic for $c = -n$ where $n \geq 0$ is an integer.
This occurs for \\ $\omega = i \alpha_{b} (n+1)$. 
From (\ref{g}), the poles of $g$ are given by the poles of $Q_{1}/Q_{2}$,
and therefore occur when $c_{1} = -n$. The corresponding frequencies
$\omega = i \alpha_{c} (n+1)$. Thus non-analyticity of $f$ and $g$ are 
as expected from the violations of (\ref{potcond}). 
This non-analyticity implies from the results in \cite{ching} that QNMs do
not completely describe the time behaviour of the scalar field at intermediate
times for the potential $V$. Since we have chosen our potential 
to match the SdS potential asymptotically, we expect the same result for the SdS
potential. The numerical results for a field in 
the SdS potential suggest power law behaviour in time for certain intermediate times.
It would be interesting to see if such a power law behaviour could be derived in our
model and the parameters the exponent would depend on. In particular, it would
be useful if in our model, we could completely obtain the time
dependence of the field and the times at which the power law behaviour
sets in. This would be a valuable pointer to the time decay of the 
field in the SdS spacetime.

\subsection{ Angular mode dependence of late time behaviour}
We now compare the late time behaviour in our model with numerical results.
The late time behaviour in the SdS spacetime was first studied in \cite{bckl}
and later in \cite{bclp}. The numerical results consider (as for the early times)
a scalar field wave function with certain initial conditions whose time
evolution is determined by numerically integrating the Klein-Gordon
equation.  However, as it is required to access late times, it is no longer
possible to work with the choice of black hole parameters used for early
times, and the authors in \cite{bclp} use $r_{b}=1$ and $r_{c}=100$ - but
now the two horizons are not so well-separated. They then study the {\em leading}
late time behaviour for the angular modes $L=0, 1, 2$. For $L=0$, they find that the 
field settles to a constant value. For $L=1,2$, the field decays exponentially
with the argument of the exponent consistent with the formula $-\alpha_{c} L$.
However, it is not possible for them to verify this proposed
formula for the higher angular modes ($L >2$) as the numerical integration 
becomes noisy. Thus the late time behaviour (being exponential decay)
is a typical QNM behaviour where the QNM is pure imaginary.
However, the numerical results display only the leading behaviour, given
by the lowest QNM which seems to be $\omega = i\alpha_{c} L$. It must be noted
that no numerical results are available on the nature of the higher order modes for a
given $L$ (although the problem is addressed by the authors of \cite{bclp} in 
an analytical approximation where the central black hole in the SdS spacetime is
disregarded). 

We consider the QNMs describing late time behaviour in our model given by
(\ref{omegaset2}). These QNMs are
proportional to $\alpha_c$ and have a non-zero real part which depends on $L$.
The dependence of the QNMs on the angular 
mode $L$ is through the maximum of the potential $V_m$
and therefore affects only their real part. 
This is very similar to 
results obtained for the nearly extreme SdS black hole in \cite{moss, lemos}.
However it is at variance to the numerical results in \cite{bclp} where at least
for the lowest QNM, the real part is zero, and the imaginary part depends on 
$L$ (although both the model QNMs and
the numerical results show a dependence on $\alpha_c$). 
Unfortunately, we also have no numerical results available on higher order
modes (for a given $L$) to compare with the higher order model QNMs.

Our approximate potential $V$ is the simplest model
which is {\em solvable} in the limit of separation of scales, and reveals the 
presence of two sets of QNMs coming into play at intermediate and
late times as seen in numerical studies. However, it does need
to be refined further to also reproduce the $L$ dependence of the 
second set of modes.
However, the refinement has to be such that the model remains tractable
at least in the physical regime of interest. This is a severe restriction.
This is because much of our success above in reproducing the striking
qualitative features has to do with correctly reproducing the asymptotic
behaviour of the SdS potential at both horizons. The SdS potential falls
off exponentially at both horizons, but with different arguments. This is
difficult to reproduce in an exactly solvable potential, and therefore,
one must piece together two potentials as we have done at a point, 
which could be the maximum of the potential. However, in this case, 
matching conditions may make the actual equation for QNMs complicated. 
The $L$ dependence seen in numerical studies for the second set of modes
and the failure of this model to reproduce them is related to the failure
of a Poschl-Teller potential to describe the $L$
dependence of the vacuum de Sitter potential. One obvious modification
we could do to our model to reproduce the $L$ dependence correctly
is to use the same Poschl-Teller potential for 
$r_b < r < r_m$ and the vacuum de Sitter potential for $r_m < r <r_c$. 
However, in that case, the matching conditions at $r_m$ become very 
complicated.  
This makes the equation for QNMs complicated, and it is no longer
possible to see the expected qualitative features emerge neatly in the 
$M\ll l$ approximation. Refinement of this model should therefore be 
concentrated on finding a tractable potential that is a good approximation
to the vacuum de Sitter potential and could be pieced to the 
Poschl-Teller potential at the maximum. It should be however remembered
here that piecing at the maximum, and resulting matching conditions
may result in absence of certain QNMs from the spectrum.
\subsection{Further investigation of the model}
We have outlined above the steps to be taken for making the QNMs of this
model agree well with numerical studies for late time behaviour of the
field. However, it must be noted that {\em no} analytical approximation
proposed so far for time behaviour of fields even in the Schwarzschild black
hole reproduces its entire time behaviour from intermediate to 
late times. The analytical model in
\cite{ferrari} reproduces the QNMs, but not the characteristic late time
tail for a Schwarzschild perturbation. It is to be noted that our model
shows exponential late time decay for the field as seen for the SdS
black hole, although the exponent
does not match numerical studies. It also shows the QNMs at intermediate
times. Therefore it offers an interesting platform to address the following
questions:

1) Can one map the complete time behaviour of a perturbation in this model,
starting from intermediate times? The incompleteness of the QNMs of this model suggests
interesting temporal behaviour when the field switches from a decay governed
by the first QNM set to the second. It would be most exciting if in this model,
a power law decay were found at these intermediate times. It would then 
shed light on similar behaviour in the SdS black hole.

2) What happens to the near-extremal limit in this model? The discrepancies
of our model QNMs with numerical studies are mainly in the angular mode 
dependence. We therefore expect that the results of taking the extremal 
limit in our model (two horizons approaching each other) should be
a good approximation to the perturbation behaviour when the SdS black 
hole approaches the Nariai black hole. Investigations of the near-extremal limit
for asymptotically flat black holes (Reissner-Nordstrom) show 
pecularities \cite{andersson}. Our model could be investigated for such
pecularities in the near-extremal limit of the SdS black hole. 

It is possible to address the above two questions as this model appears tractable.
For reasons elaborated above, such an investigation would indeed be valuable
in understanding the response of the SdS black hole to perturbations.
This investigation, and further refinements of this model to match numerical 
studies are the subjects of future work.  

\section{Acknowledgement}
The author would like to thank Eric Poisson for useful comments.
This work was supported by a fellowship of the Alexander von Humboldt Foundation.


\begin{thebibliography}{100}
\bibitem{vish} C.V. Vishveshvara, Nature {\bf 227} (1970) 936.
\bibitem{nollert} H. Nollert, Class. Quant. Grav. {\bf 16} (1999) R159.
\bibitem{kokkotas} K.D. Kokkotas and B.G. Schmidt, Living Rev. Rel. {\bf 2}\\
(www.livingreviews.org/Articles/Volume2/1999-2kokkotas).
\bibitem{mellor} F. Mellor and I. Moss, Phys. Rev. {\bf D41} (1990) 403.
\bibitem{moss} I. Moss and J.P. Norman, Class. Quant. Grav. {\bf 19} 
(2002) 2323.
\bibitem{lemos} V. Cardoso and J.P.S. Lemos, gr-qc/0301078.
\bibitem{bckl} P.R. Brady, C. Chambers, W. Krivan and P. Laguna, 
Phys. Rev. {\bf D55} (1997) 7538. 
\bibitem{bclp} P.R. Brady, C. Chambers, W. Laarakkers and E. Poisson,
Phys. Rev. {\bf D 60} (1999) 064003.
\bibitem{ferrari} V. Ferrari and B. Mashhoon, Phys. Rev. {\bf D30} (1984) 295.
\bibitem{mann1} J.S.F Chan and R.B. Mann, Phys. Rev. {\bf D55} (1997) 7546. QNMs for the
BTZ black hole are studied analytically here, but using an approximate potential 
that reproduces the behaviour of the BTZ potential near the horizon.
\bibitem{gs} T.R. Govindarajan and V. Suneeta, Class. Quant. Grav. 
{\bf 18} (2001) 265. 
\bibitem{lemos1} V. Cardoso and J.P.S. Lemos, Phys. Rev. {\bf D63} (2001) 124015;
Phys. Rev. {\bf D64} (2001) 084017.
\bibitem{hh} G.T. Horowitz and V.E. Hubeny, Phys. Rev. {\bf D62} (2000) 024027.
\bibitem{mann2} J.S.F. Chan and R.B. Mann, Phys. Rev. {\bf D59} (1999) 064025.
\bibitem{mann3} B. Wang, E. Abdalla and R.B. Mann, Phys. Rev. {\bf D65} (2002)
084006.
\bibitem{zan} R. Aros, C. Martinez, R. Troncoso and J. Zanelli, Phys. Rev. {\bf D67},
(2003) 044014.
\bibitem{bss} D. Birmingham, I. Sachs and S.N. Solodukhin, 
Phys. Rev. Lett. {\bf 88} (2002) 151301.
\bibitem{son} D.T. Son and A.O. Starinets, JHEP {\bf 0209}(2002)042.
\bibitem{strominger} A. Strominger, JHEP {\bf 0110}(2001)034;
M. Spradlin, A. Strominger and A. Volovich, {\em Les Houches
lectures on deSitter space}, hep-th/0110007 and references therein.
\bibitem{branco} E. Abdalla, K.H.C. Castello-Branco and A. Lima-Santos, 
Phys. Rev. {\bf D66} (2002) 104018.
\bibitem{ching} E.S.C. Ching, P.T. Leung, W.M. Suen and K. Young, Phys. Rev. {\bf D54} (1996) 3778.
\bibitem{andersson} N. Andersson and H. Onozawa, Phys. Rev. {\bf D54} (1996) 7470.
\end{thebibliography}
\end{document}